\documentclass{iau} 
\usepackage{graphicx}

\title[The interaction of CCSN ejecta with a companion] 
{The interaction of core-collapse supernova ejecta with a stellar companion}

\author[Zheng-Wei Liu et al.]   
{Zheng-Wei Liu$^{1,2}$,
T. M. Tauris$^{3}$, F. K. R\"opke$^{4,5}$, T. J. Moriya$^{6}$, M. Kruckow$^{1,2}$, R. J. Stancliffe$^{3}$, R. G. Izzard$^{7}$}

\affiliation{$^1$Yunnan Observatories, Key Laboratory for the Structure and Evolution of Celestial Objects, CAS, Kunming 650216, China, $^{2}$Center for Astronomical Mega-Science, CAS, Beijing, China \\email: {\tt zwliu@ynao.ac.cn}

$^{3}$Argelander-Institut f\"ur Astronomie, Auf dem H\"ugel 71, D-53121 Bonn, $^{4}$Heidelberger Institut f\"ur Theoretische Studien, Schloss-Wolfsbrunnenweg 35, D-69118 Heidelberg, Germany, $^{5}$Zentrum f\"ur Astronomie der Universit\"at Heidelberg, Institut f\"ur Theoretische Astrophysik, Philosophenweg 12, D-69120 Heidelberg, Germany, $^{6}$National Astronomical Observatory of Japan, $^{7}$University of Surrey, Guildford, Surrey GU2 7XH, United Kingdom. }

\pubyear{2018}
\volume{xxx}  
\jname{IAU Symposium 346: High-mass X-ray binaries: illuminating the passage from massive binaries to merging compact objects}
\editors{A.C. Editor, B.D. Editor \& C.E. Editor, eds.}
\begin{document}

\maketitle

\begin{abstract}
The progenitors of many core-collapse supernovae (CCSNe) are expected to be in binary systems. By performing a series of three-dimensional hydrodynamical simulations, we investigate how CCSN explosions affect their binary companion. We find that the amount of removed stellar mass, the resulting impact velocity, and the chemical contamination of the companion that results from the impact of the SN ejecta, strongly increases with decreasing binary separation and increasing explosion energy. Also, it is foud that the impact effects of CCSN ejecta on the structure of main-sequence (MS) companions, and thus their long term post-explosion evolution, is in general not be dramatic.

\end{abstract}

\firstsection 
\section{Introduction}

\begin{figure}
\begin{center}
 \includegraphics[width=4.5in]{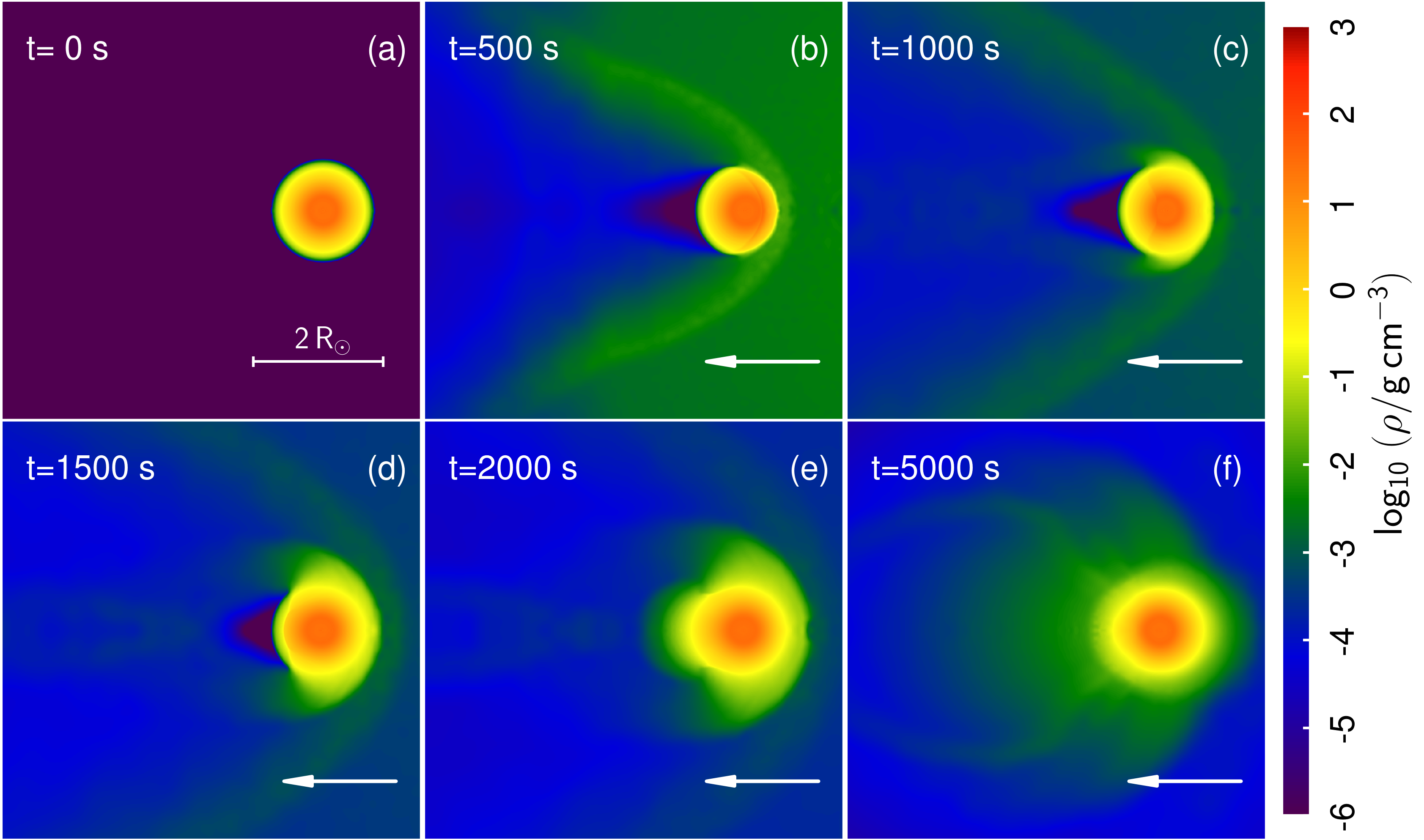} 
 \caption{Density distributions of all gas material as a function of the time in our impact simulations for a G/K-dwarf 
          companion model with a binary separation of $5.48\,R_{\odot}$. The direction of motion of the incoming SN shell front is 
          from right to left (see arrow symbols). The color scale shows the logarithm of the mass density in $\rm{g\,cm^{-3}}$.}
   \label{fig1}
\end{center}
\end{figure}

The discovery of many low-mass X-ray binaries and millisecond pulsars in tight orbits, i.e. binary neutron stars with orbital periods of less than a few hours, provides evidence for supernova (SN) explosions in close binaries with low-mass companions. The nature of the SN explosion determines whether any given binary system remains bound or is disrupted (\cite[Hills 1983]{Hills1983}). An additional consequence of the SN explosion is that the companion star is affected by the impact of the shell debris ejected from the exploding star (\cite[Wheeler et al. 1975]{Whee1975}). Besides chemical enrichment, such an impact has kinematic effects and may induce significant mass loss and heating of the companion star. Core-collapse supernovae (CCSNe) arise from massive stars. There is growing observational evidence that the fraction of massive stars in close binary systems is large. \cite[Sana et al. (2012)]{Sana2012} found that more than $70\%$ of massive stars are in close binary systems, which supports the idea that binary progenitors contribute significantly to the observed CCSNe.

After a SN explosion occurs in a binary system, the ejected debris is expected to expand freely and eventually impact the companion star. The companion star may be significantly heated and shocked by the SN impact, causing the envelope of the companion star to be partially removed due to the stripping and ablation mechanism (e.g., \cite[Wheeler et al. 1975]{Whee75}, \cite[Marietta et al. 2000]{Marietta2000}, \cite[Liu et al. 2012]{Liu2012}, \cite[Liu et al. 2013]{Liu2013}, \cite[Pan et al. 2012]{Pan2012}, \cite[Hirai et al. 2018]{Hirai2018}). In this work, we perform impact simulations using a three-dimensional (3D) smoothed particle hydrodynamics (SPH) method to systematically study, for the first time, the impact of CCSN ejecta on MS companion stars.

\begin{figure}
\begin{center}
 \includegraphics[width=2.6in]{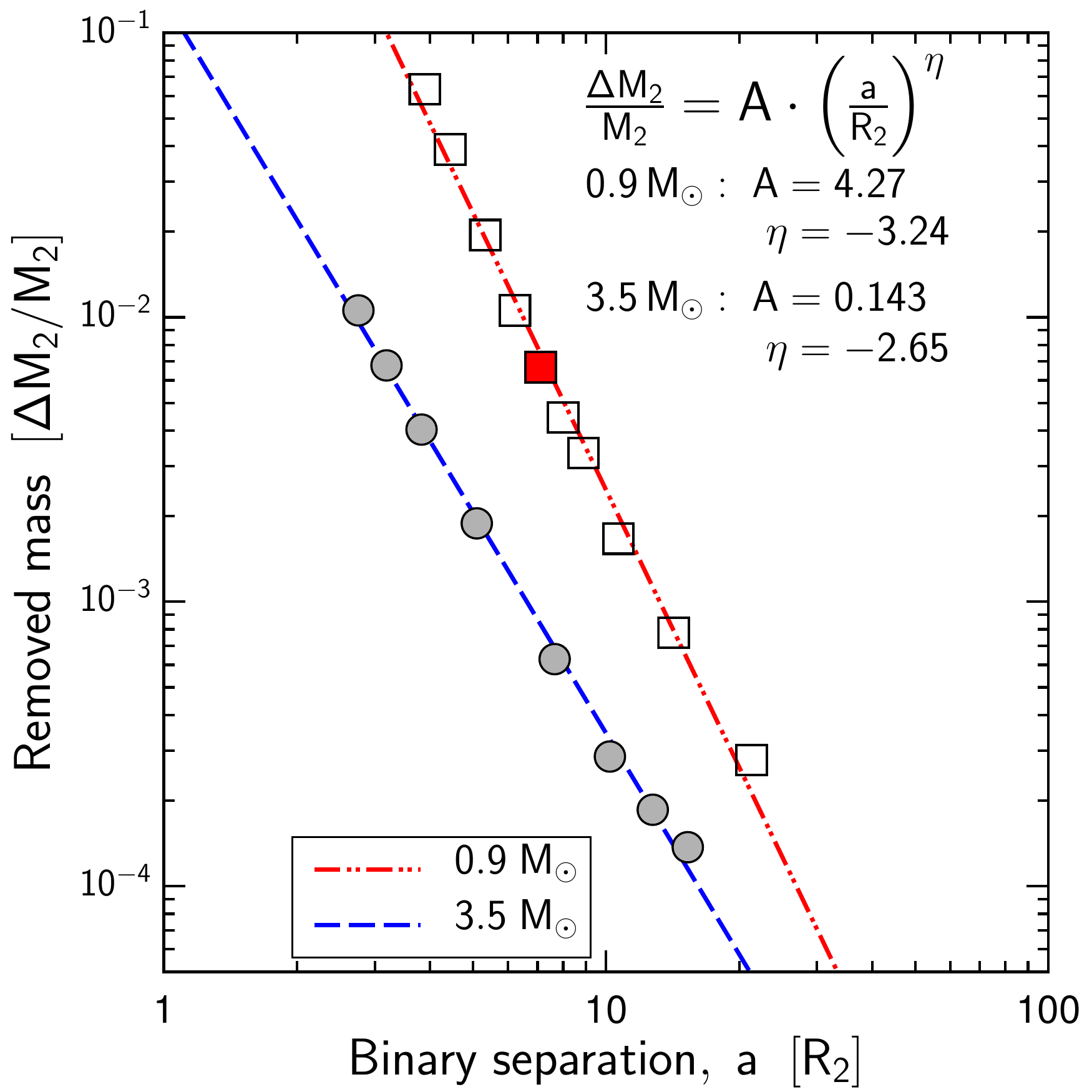} 
 \includegraphics[width=2.6in]{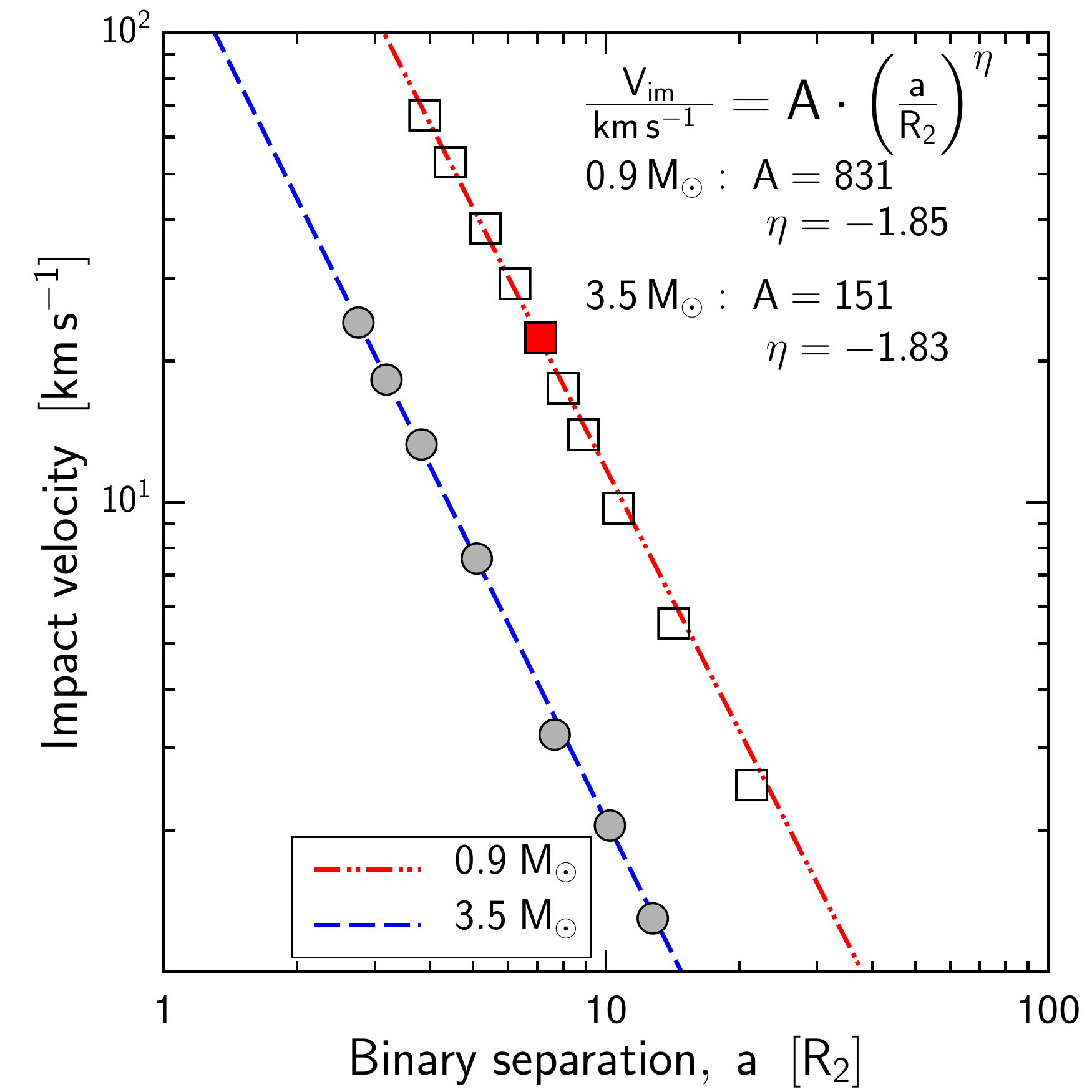} 
 \includegraphics[width=2.8in]{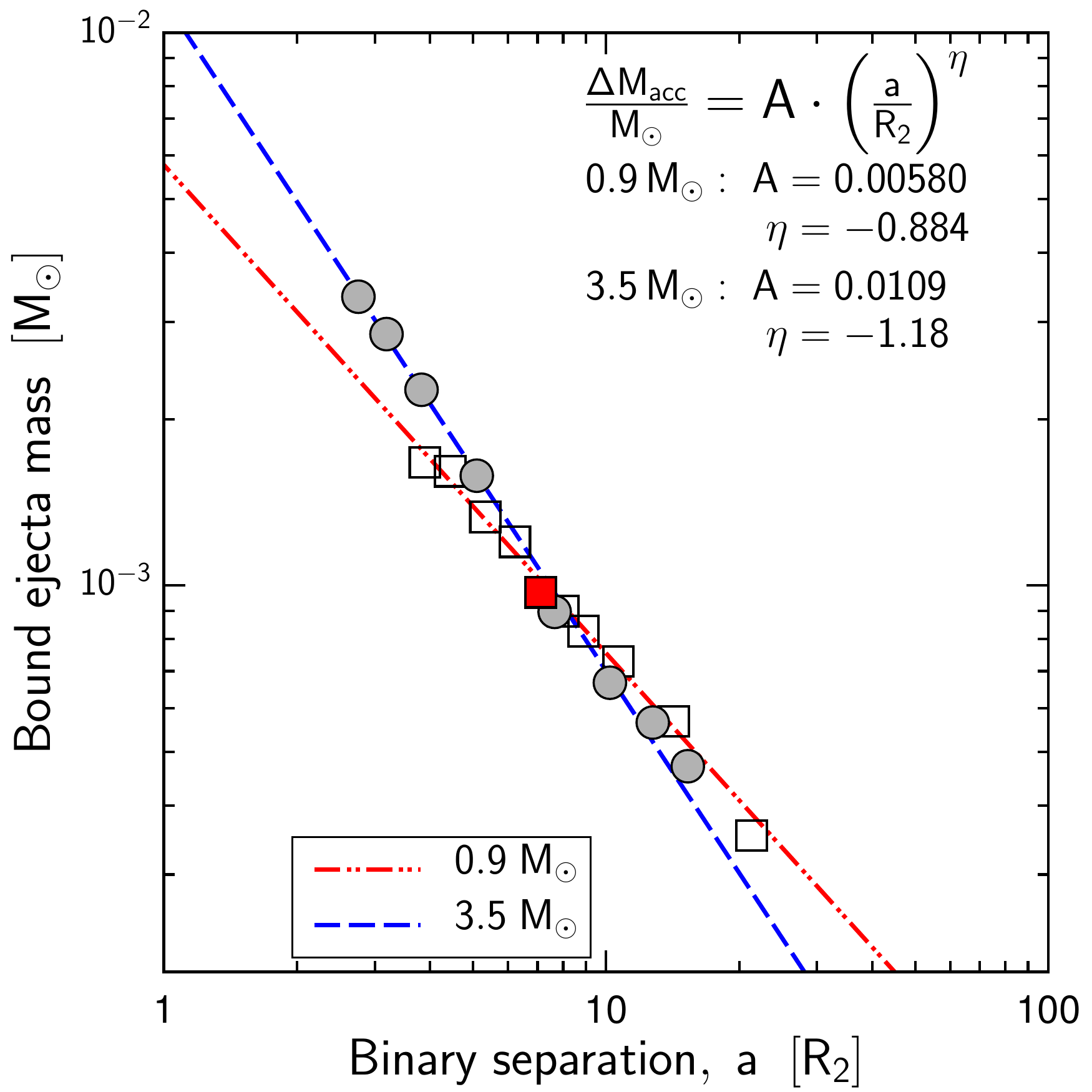} 
 \caption{Total removed companion mass (top-left panel), resulting impact velocity of the companion star (top-right panel) and the amount of accreted contamination
           from the SN ejecta (bottom panel), as a function of initial binary separations for a G/K-dwarf (square symbols, $M_{2}=0.9\,M_{\odot}$ and $R_2\approx0.77\,R_{\odot}$) and 
           a late-type B-star (filled circle symbols, $M_{2}=3.5\,M_{\odot}$ and $R_2\approx2.18\,R_{\odot}$) companion model. Power-law fits
           are also  presented in each panel.}
   \label{fig2}
\end{center}
\end{figure}

\section{Results and Conclusions}

We use the {\sc BEC} stellar evolution code to construct the detailed companion structure at the moment of SN explosion. The impact of the SN blast wave on the companion star is followed by means of 3D SPH  simulations using the {\sc Stellar GADGET} code (\cite[Pakmor et al. 2012]{Pakmor2012}). Figure~\ref{fig1} illustrates the temporal density evolution of the SN ejecta and companion material of our hydrodynamics simulations for a G/K-dwarf companion model. Figure~\ref{fig2} shows the effects of varying the orbital separation parameter $a/R_{2}$, by a factor of about 6, on the total amount of removed companion mass ($\Delta M_2$), the resulting impact velocity ($v_{\rm im}$), and total accumulated ejecta mass ($\Delta M_{\rm acc}$), for the $0.9\,M_{\odot}$ and $3.5\,M_{\odot}$ companion star models.

To discuss the effects of an explosion on the companion star in CCSNe of Type~Ib/c, we perform populations synthesis calculations for SNe with the {\tt binary\_c/nucsyn} code (\cite[Izzard et al. 2004]{Izzard2004}, \cite[Izzard et al. 2009]{Izzard2009}). Under an assumption of a Galactic star-formation rate of $0.68-1.45\,M_{\odot}\,\rm{yr^{-1}}$ and the average stellar mass of the Kroupa initial mass function ($0.83\,M_{\odot}$), the total CCSN rate of the Galaxy predicted from their population synthesis is $0.93$--$1.99\times10^{-2}\,\rm{yr^{-1}}$ (36\% SNe~Ib/c, 10\% SNe~IIb, 54\% SNe~II), consistent with the estimated Galactic CCSN rate of $2.30\pm0.48\times10^{-2}\,\rm{yr^{-1}}$ from recent surveys (\cite[Li et al. 2011]{Li2011}).

\begin{figure}
\begin{center}
 \includegraphics[width=4.5in]{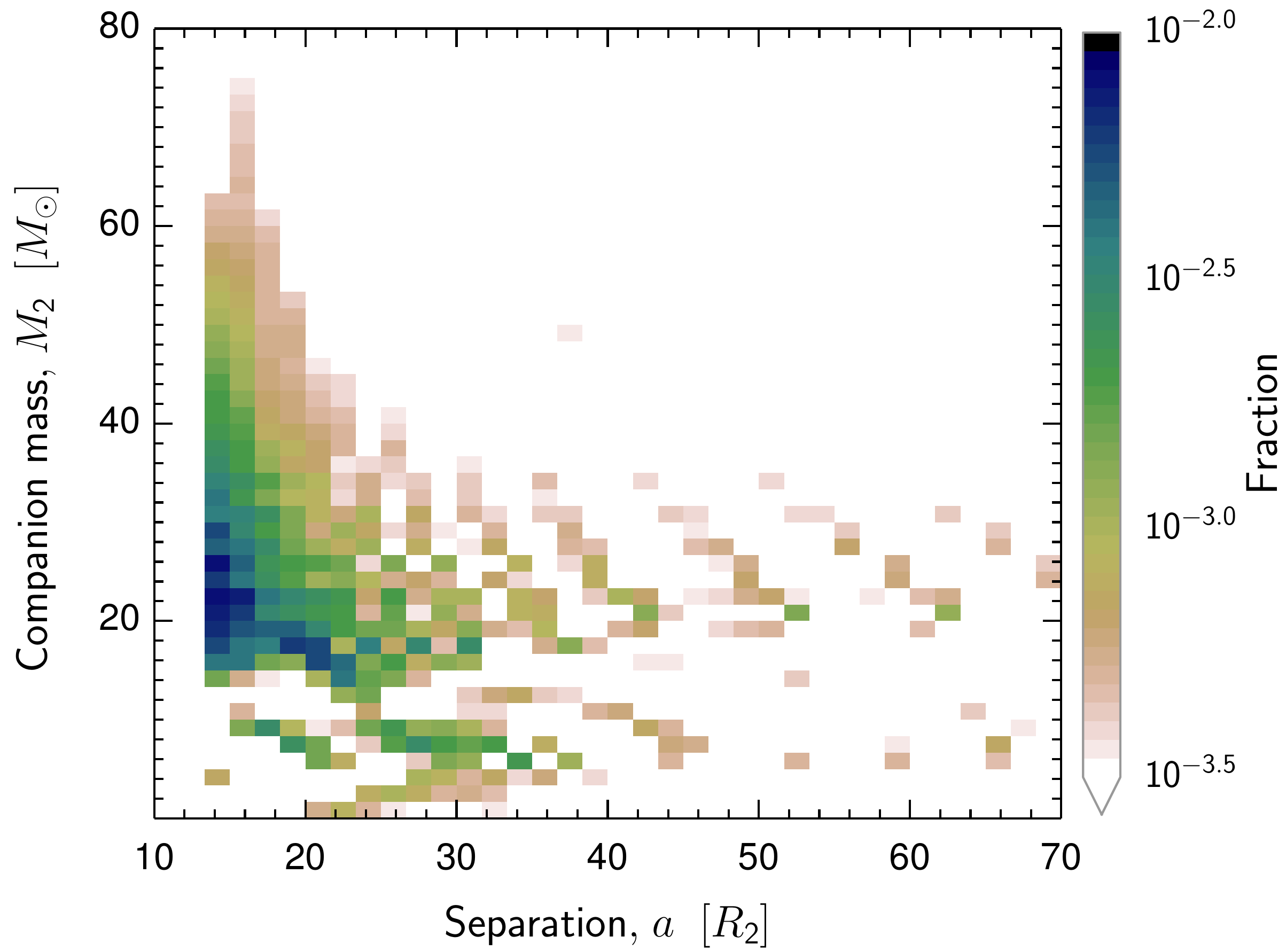} 
 \caption{Population synthesis distribution of the companion star mass ($M_{2}$) as a function of the binary separation ($a$) in 
          case the SN explodes as a Type Ib/c. Nothing is plotted in the regions with number fraction smaller than $10^{-3.5}$.}
   \label{fig3}
\end{center}
\end{figure}

In our populations synthesis calculations, most SNe~Ib/c have an orbital separation of $\gtrsim5.0\;R_2$, which is about a fraction of $\gtrsim95\%$ in our binary population synthesis calculations. Furthermore, with the distributions of $a/R_{2}$ in Fig.~\ref{fig3}, we can simply estimate $\Delta M_2$, $v_{\rm im}$ and $\Delta M_{\rm acc}$ by applying our power-law relationships stated in Fig.~\ref{fig2}. We caution that there are large uncertainties in population synthesis studies, which may influence the results. In this case, these mainly relate to the input physics of common-envelope evolution and the subsequent Case BB roche-lobe overflow from the naked helium star prior to its explosion.

We have investigated the impact of SN ejecta on the companion stars in CCSNe of Type~Ib/c using the SPH code {\sc Stellar GADGET}. Our main results can be summarized as follows (see also \cite[Liu et al. 2015]{Liu2015}): 

\begin{itemize}

\item[\,i)] The dependence of total removed mass ($\Delta M_2$), impact velocity ($v_{\rm im}$) and the amount of accreted SN ejecta mass ($\Delta M_{\rm acc}$) on the pre-SN binary separation ($a$) can be fitted with power-law functions. All three quantities are shown to decrease significantly with increasing a, as expected (see Fig.~\ref{fig2}). \\[-2.0ex]
\item[\,ii)] If our population synthesis is correct, we predict that in most CCSNe less than $5\%$ of 
           the MS companion mass can be removed by the SN impact 
           (i.e. $\Delta M_2/M_2<0.05$). In addition, the companion star typically receives an impact velocity, $v_{\rm im}$,
           of a~few $10\,\rm{km\,s^{-1}}$, and the amount of SN ejecta captured by the companion star after the explosion, $\Delta M_{\rm acc}$, 
           is most often less than $10^{-3}\,M_{\odot}$. \\[-2.0ex]
\item[\,iii)] Because a typical CCSN binary companion is relatively massive and can be located at a large pre-SN distance,
           we do not expect, in general, that the effects of the SN explosion on the post-impact stellar evolution will be very dramatic. \\[-2.0ex]
\item[\,iv)] In the closest pre-SN systems, the MS companion stars are affected more strongly by the SN ejecta impact, leading to
           $\Delta M_2/M_2\simeq 0.10$, $v_{\rm im}\simeq 100\;{\rm km\,s}^{-1}$ and $\Delta M_{\rm acc}\simeq 4\times 10^{-3}\;M_{\odot}$, depending
           on the mass of the companion star. In addition, these stars are significantly bloated as a consequence of internal heating by
           the passing shock wave. \\[-2.0ex]
\item[\,v)] It is possible that the SN-induced high velocity stars (HVSs), or more ordinary, less fast, runaway stars, 
           may be contaminated sufficiently to be identified
           by their chemical peculiarity as former companion stars to an exploding star if mixing processes are not 
           efficient on a long timescale.
\end{itemize}

\end{document}